\documentstyle[12pt]{article}
\begin{document}
\begin{titlepage}
\title{{\bf Scalar--Tensor Theories of Gravity with
$\Phi$--dependent masses}
\thanks{Work partially supported by CICYT under
contract AEN90--0139.}}

\author{
{\bf J.A. Casas},\ \
{\bf J. Garc\'{\i}a--Bellido}\thanks{Supported by MEC--FPI Grant.
e--mail: bellido@iem.csic.es}\ \
and \  {\bf M. Quir\'os}\thanks{
e--mail: quiros@cernvm.bitnet} \\
Instituto de Estructura de la Materia. \ CSIC\\
Serrano, 123\ \ \ E--28006\ \ Madrid.\ \ Spain}

\date{}
\maketitle
\vspace{1.5cm}
\def\baselinestretch{1.15}
\begin{abstract}
We study new physical phenomena and
constraints in generalized scalar--tensor theories of gravity
with $\Phi$--dependent masses. We investigate a scenario (which
can arise in string theories) with two types of $\Phi$--dependent
masses which could correspond to visible and dark matter
sectors. The parameters of this theory are constrained from
post--Newtonian bounds, primordial nucleosynthesis and the age
of the Universe. We present a perfect fluid formalism for the dark
matter sector with variable masses and find an entropy increase
effect during the matter era and, in principle, a measurable
effect on the motion of the halo of spiral galaxies.
For the case of string effective theories, the constancy
of gauge couplings provide new bounds which are orders of
magnitude stronger than the previous ones.
\end{abstract}

\vskip-17.5cm
\rightline{{\bf IEM--FT--40/91}}
\vskip3in

\end{titlepage}

\newpage
\def\baselinestretch{1.25}

\section{Introduction}

General Relativity (GR) is the simplest classical field theory for
a metric tensor $g_{\mu\nu}$ with general coordinate transformation
invariance \cite{WEI}. It correctly describes the behaviour of
particles with constant masses under a gravitational field and
constitutes a beautiful geometric generalization of Newton's
theory of gravity \cite{SCT}. It is based on the action
\begin{equation}
\label{GR}
S = \int d^4x \sqrt{-g} \left(\frac{1}{G} R + 16\pi{\cal L}_m
\right)
\end{equation}
where $G=M_{Pl}^{-2}$, $R$ is the curvature scalar and
${\cal L}_m$ the usual matter
Lagrangian. In spite of its simplicity
and phenomenological success in different
contexts (particularly in the cosmological and post--Newtonian
scenarios), several alternative theories of gravitation have
appeared since Einstein's proposal \footnote{It is worth
recalling that GR is unable to explain some crucial aspects,
like the value of the cosmological constant, the universe initial
condition problem, the existence of classical singularities and its
connection with quantum mechanics.}. The simplest extension
of GR is the  Jordan--Brans--Dicke (JBD) theory of gravity
\cite{JBD}, based on the action
\begin{equation}
\label{JBD}
S = \int d^4x \sqrt{-g} \left(\Phi R
- \frac{\omega}{\Phi} g^{\mu\nu}\partial_\mu\Phi \partial_\nu\Phi
+ 16\pi {\cal L}_m \right)
\end{equation}
where $\Phi$ is a scalar field with dimensions of mass--squared
and $\omega$ a constant
\footnote{GR is recovered in the limit $|\omega|\rightarrow\infty$.
Post--Newtonian \cite{SCT,VIK} and nucleosynthesis \cite{FT26}
constraints impose $|\omega| > 250$\ \ $(2\sigma$ limit).}.
These theories have recently recovered great interest since they
have been proposed as the arena for 'extended inflation'
\cite{EI,MQ}, a new inflationary scenario which has the potential
to overcome some of the traditional problems of previous
schemes. A first generalization of JBD theories, which has been
widely studied in the literature, are the so called
scalar--tensor (ST) theories, based on the action (\ref{JBD})
with a variable parameter $\omega(\Phi)$ \cite{SCT}.

In this paper we analyze the thermodynamical properties and
phenomenological constraints of a different generalization
of JBD theories, previously considered in
refs.[8--11], in which the scalar field is coupled
to matter \footnote{During the radiation dominated era these
couplings are irrelevant since the kinetic terms dominate the
matter terms; as a consequence, the theory behaves like GR with
constant couplings \cite{FT27}.}. In particular, it is specially
relevant the case of matter fields with $\Phi$--dependent
masses $m_n(\Phi)$.

There is a very strong and fundamental reason for considering
theories with $\Phi$--dependent masses. Namely, low--energy effective
theories of this type are those that emerge naturally from
superstring theory \cite{GSW}, the only known consistent theory
unifying all the fundamental forces. In this case,
the scalar field $\Phi$ is related to the dilaton and moduli
fields (scalar fields in the graviton multiplet), and its
coupling with matter is determined by the string dynamics and the
compactification mechanism. Before supersymmetry
breaking, there are flat directions (corresponding to massless
fields) and fields acquiring vacuum expectation values
(non--flat directions), corresponding to spontaneous gauge
symmetry breaking ({\em e.g.} via Fayet--Iliopoulos terms \cite{FI}),
leading to $\Phi$--dependent masses. When supersymmetry is broken
({\em e.g.} by gaugino condensation \cite{GAU}) all scalar fields
acquire a mass term proportional to $m_{3/2}^2$ (which is
$\Phi$--dependent), directly at tree level or communicated via
radiative corrections. The former flat and
non--flat directions may get different $\Phi$--dependent masses
through the supersymmetry breaking mechanism.

Two gravity theories which differ by a conformal redefinition of
the metric can be shown to be equivalent
(see e.g. refs.\cite{DGG,FT27,KKO})\footnote{The equivalence
between conformal frames for physical
observables can be understood from the fact that they
are constructed as dimensionless variables.}.
As a consequence, scalar--tensor theories in which the masses
of all particles have the {\em same} $\Phi$--dependence can be
put in the form of standard JBD theory through a conformal
redefinition of the metric and the scalar field. However, this
is not the most general case in effective theories of
gravity ({\em e.g.} from superstrings), where {\em different}
$\Phi$--dependent masses can appear.
Damour {\em et al.} \cite{DGG} considered a theory with two
matter sectors (visible and dark matter) with different
$\Phi$--dependences. They analyzed the cosmological behaviour of
the theory and imposed some constraints on its parameters from
the radar time--delay measurements, the age of the universe and
the time variation of $G$ .
Holman {\em et al.} \cite{HKW} considered the same theory in the
context of extended inflation (see also \cite{FT37}) and
constrained its parameters from the isotropy of the cosmic
background radiation.

In this paper we analyze new physical phenomena and constraints
associated with this theory.
In particular, we present a perfect fluid formalism for
dark matter with variable masses and find a big entropy
production during the matter era. We analyze the
post--Newtonian (PPN) scenario for such a theory and find an
effect on the motion of the dark matter halo of
spiral galaxies. We also give updated constraints on the
parameters of the theory from primordial nucleosynthesis,
using recent data.
Finally, we consider the possibility that this kind of theory
could emerge from superstrings (as it was mentioned above this
is quite natural and it has been previously suggested in the
literature [8--11]). In that case
there is also a dilaton coupling to the electromagnetic
sector. We then derive new very strong constraints on the
parameters from the constancy of the electromagnetic coupling.

In sect.2 we consider a gravity theory with two different matter
sectors, which could arise from superstrings.
In sect.3 we study new
physical phenomena arising from this theory. In
particular, we find that particles with variable masses do not
follow geodesics but satisfy Newton's second law. There is also an
entropy increase during the matter era, as a consequence of the
non--conservation of the energy--momentum tensor and, in
principle, a measurable effect on the motion of the halo of
spiral galaxies. In sect.4 we constrain the
parameters of this theory from primordial nucleosynthesis.
In sect.5 we find very strong bounds on $\omega$ in the context of string
theories from the variation of the gauge couplings.
In sect.6 we draw our conclusions and outlook.

\section{Gravity theories with different $\Phi$--dependent masses}

In this section we consider the simplest theory of gravity with
different $\Phi$--dependent masses, the case with only two
different matter sectors: one associated with 'dark' matter
(invisible sector) and the other being the usual observed
baryonic matter (visible sector). This kind of scenario has been
previously considered by Damour {\em et al.} in Ref. \cite{DGG}.
The action can be written in the Einstein frame
as a two parameter theory \footnote{From here on we will work in
units of Planck's mass.}
\begin{equation}
\label{SMD}
S = \int d^4x \sqrt{-g} \left(R - \frac{1}{2} (\partial\phi)^2
+ 16\pi e^{\beta_I\phi} {\cal L}_{m_I}^{(0)}
+ 16\pi e^{\beta_V\phi} {\cal L}_{m_V}^{(0)}  \right) \ \ ,
\end{equation}
where the scalar field $\phi$ corresponds to twice the dilaton
field in string effective theories, $(\beta_I,\beta_V)$
parametrize the invisible ({\em i.e.} dark) and visible ({\em
i.e.} baryonic) matter sectors\footnote{We restrict ourselves,
for simplicity, to the case $\ \beta_I, \beta_V>0$. Note that
our notation differs from that of ref.\cite{DGG} by the
definition of the dilaton, $\sigma\rightarrow\frac{1}
{\sqrt{2}}\phi$.} and the
matter Lagrangian can be written as ${\cal L}_{m_j}(\phi) =
e^{\beta_j\phi} {\cal L}_{m_j}^{(0)},\;j=I,V$, where, suppressing
the $j$ index \cite{WEI},
\begin{equation}
\label{LM}
{\cal L}_m^{(0)} =  \frac{-1}{\sqrt{-g}} \sum_n\ m_n^{(0)}
\int^{\infty}_{-\infty} d\tau_n \left[ -g_{\mu\nu}(x_n)
\frac{dx^{\mu}_n}{d\tau_n}\frac{dx^{\nu}_n}{d\tau_n}
\right]^{1/2} \delta^{(4)}(x-x_n)
\end{equation}
and $n$ labels a set of classical point particles
with variable masses $m_n(\phi) = e^{\beta\phi} m_n^{(0)}$
\footnote{The description of dust particles with variable masses
through this Lagrangian has been previously considered in
refs.\cite{DGG,FT27}.}.
We can then perform a conformal redefinition of the metric in order
to leave constant masses for visible matter (we call this the
physical frame \cite{FT27}) \footnote{We know
that physics cannot distinguish between conformal frames,
therefore we are free to choose whatever masses are made constant.
We choose constant visible masses for convenience since
they give constant units of measure and therefore visible
particles follow trajectories which are geodesics of the metric.}.
The resulting theory has the form of a
generalized JBD theory with variable masses for the dark matter
sector
\begin{equation}
\label{SBD}
S = \int d^4x \sqrt{-g} \left(\Phi R -
\frac{\omega}{\Phi} (\partial\Phi)^2
+ 16\pi\Phi^{\sigma}{\cal L}_{m_I}^{(0)}
+ 16\pi{\cal L}_{m_V}^{(0)} \right)
\end{equation}
where $\ \Phi=e^{-2\beta_V \phi}\ $, $m_n(\Phi) = \Phi^\sigma m_n^{(0)}$
and the two parameters
$(\omega,\sigma)$ are given by
\begin{equation}
\label{OSB}
\begin{array}{rl}
2\omega+3&=
{\displaystyle
\frac{1}{4\beta^2_V} }\vspace{2mm}\\
1-2\sigma&=
{\displaystyle
\frac{\beta_I}{\beta_V} }  \ \ .
\end{array}
\end{equation}

If we assume that the universe evolution (during the matter era)
is {\em dominated by dark matter}, the equations of motion
corresponding to the theory (\ref{SBD}) are given by
\begin{equation}
\label{SEQ1}
R_{\mu \nu}
=\frac{8\pi}{\Phi} \left( \frac{1}{2}g_{\mu \nu}
T^\lambda_{\ \lambda} - T_{\mu \nu} \right) -\frac{\omega}
{\Phi^2} \partial_{\mu}\Phi
\partial_{\nu}\Phi-\frac{1}{\Phi}\left(D_{\mu}D_{\nu}
\Phi+\frac{1}{2}g_{\mu \nu} D^2 \Phi \right)
\end{equation}
\begin{equation}
\label{SEQ2}
(2\omega+3) D^2\Phi=8\pi (1-2\sigma)
T^\lambda_{\ \lambda} \ \ ,
\end{equation}
where\footnote{It is clear from eq.(\ref{SEQ2}) that $\sigma=\frac{1}{2}$
corresponds to GR since the scalar field $\Phi$ is then constant.}
\begin{equation}
\label{TM}
T^{\mu\nu}\equiv T^{\mu\nu}_m(\Phi)
= \frac{1}{\sqrt{-g}} \sum_n \ m_n(\Phi)
\int^{\infty}_{-\infty} d\tau_n
\frac{dx^{\mu}_n}{d\tau_n}\ \frac{dx^{\nu}_n}{d\tau_n}
\ \delta^{(4)}(x-x_n)
\end{equation}
is the energy--momentum tensor in Lagrangian formalism,
satisfying $\ T^{\lambda}_{\ \lambda} = {\cal L}_{m_I}(\Phi)=\Phi^\sigma
{\cal L}_{m_I}^{(0)}$.
Making use of the Bianchi identities
$\ (R^{\mu\nu}+\frac{1}{2}g^{\mu\nu}R)_{;\nu} = 0$,
the identity $\ \Phi_{;\nu}R^{\mu\nu} = (D^2\Phi)^{;\mu}
- D^2(\Phi^{;\mu})\ $ and the equations of motion
(\ref{SEQ1}, \ref{SEQ2}), we find the energy--momentum
conservation equation
\begin{equation}
\label{EMC}
T^{\mu\nu}_{\ \ ;\nu} = \sigma \frac{\partial^{\mu}\Phi}{\Phi}
\ T^\lambda_{\ \lambda} \ \ .
\end{equation}
Therefore, it is the coupling of the JBD scalar to the
matter Lagrangian which violates the strong equivalence
principle (EP) in the physical frame, as expected. (It will be
shown in the next section that the weak EP is also violated.)
It is important to stress the fact that the non--conservation of
the energy--momentum tensor does not affect its perfect fluid
form, that is, the geometrical properties of the fluid.
The covariant non--conservation of the energy--momentum tensor
just expresses the fact that there is an
energy exchange between matter and the scalar, and not only with
the graviton. Actually, it is a dissipative effect.

We will see that the post--Newtonian and cosmological evolution
of theories described by the action (\ref{SBD}) exhibit new
physical phenomena and can be bounded by very strong
phenomenological constraints.

\section{New physical phenomena}

In this section we study the new physical phenomena
characteristic of theories described by the action (\ref{SBD}).
We will show that dark matter
particles with variable masses do not follow geodesics,
but satisfy Newton's second law.
We suggest that dark matter with variable masses might affect
the motion of the halo of spiral galaxies.
We also give a perfect fluid
formalism for dark matter with variable masses, characterized by
a non--adiabatic expansion of the universe during the matter era
(and the subsequent entropy production), as a consequence of the
non--conservation of the energy--momentum tensor.

\subsection{Particle trajectories}

The particle trajectories corresponding to the dark matter in (\ref{SBD})
are given by
\begin{equation}
\label{geo}
\frac{d^2x^{\mu}}{d\tau^2} + \Gamma^{\mu}_{\nu\lambda}
\frac{dx^{\nu}}{d\tau}\frac{dx^{\lambda}}{d\tau} +
\sigma \frac{\partial^{\mu}\Phi}{\Phi} = 0
\end{equation}
(we have suppressed the subindex $n$)
which shows that those particles do not follow geodesics
\footnote{It is important to note that equation (\ref{geo})
exactly corresponds to the geodesic of the metric $\ \tilde{g}
_{\mu\nu}(x)=\Phi(x)^{2\sigma} g_{\mu\nu}(x)\ $
since $\ \tilde{\Gamma}^{\mu}_{\nu\lambda}
=\Gamma^{\mu}_{\nu\lambda}-\frac{1}{2}
\partial^\mu (\ln\Phi^{2\sigma})\ g_{\nu\lambda}\ $.
Of course, baryonic matter do follow geodesics of $g_{\mu\nu}(x)$.},
since their trajectories are modified by the term
$\ \sigma {\displaystyle \frac{\partial^\mu \Phi}{\Phi}\  }$.
To understand its significance, consider the
Newtonian approximation. In GR, the acceleration of a
non--relativistic particle in a weak and static gravitational
field ($g_{oo}\simeq -1+\frac{2GM}{r},\; g_{ij}\simeq \delta_{ij},\;
g_{oi}\simeq 0$) is \cite{WEI}
\begin{equation}
\label{ac1}
\frac{d^2x^i}{dt^2}\simeq\ -\Gamma^i_{oo} \simeq\ \frac{1}{2}
\frac{\partial g_{oo}}{\partial x_i} \simeq\ - \frac{GM}{r^2}
\frac{x^i}{r}
\end{equation}
{\em i.e.}, the Newtonian acceleration due to the gravitational
attraction of a mass $M$. In our case, the modification of the
geodesic affects the equations of motion of such a particle
\begin{equation}
\label{ac2}
\frac{d^2x^i}{dt^2} + \sigma \frac{\dot{\Phi}}{\Phi}
\frac{dx^i}{dt} \simeq - \frac{GM}{r^2}\frac{x^i}{r}
\end{equation}
by introducing a friction force due to the variation of mass.
This corresponds exactly to Newton's second law
\begin{equation}
\label{2LN}
\frac{d}{dt} (m v^i) = F^i \ \ ,
\end{equation}
as one would expect. This is a
check of the validity of eqs.(\ref{SMD},\ref{LM}) for the classical
description of particles with variable masses.

It is interesting to note that the extra term in (\ref{geo})
violates the strong EP, since in a local inertial frame
($g_{\mu\nu} = \eta_{\mu\nu},\ \Gamma^{\lambda}_{\mu\nu} = 0$)
there is still an acceleration acting on the particles. Moreover,
it also violates the weak EP, since matter particles in the
visible and invisible sector have different
accelerations \footnote{In comparison,
JBD theory ($\sigma=0$, $\omega=$ constant)
only violates the 'very' strong EP,
since the scalar $\Phi$ is not constant in a local
inertial frame, thus changing Newton's constant $G$ from point
to point \cite{WEI}.}.

\subsection{Post--Newtonian effects:
The halo of spiral galaxies}

{}From dynamical
observations of spiral galaxies we know that there is a halo
with great amounts of 'dark' matter \cite{MKT,RST}. If
the dark matter in the halo were composed of particles with
variable masses, it would have, in principle, a measurable
effect on the dynamical motion of the halo.

The analysis of the post--Newtonian motion of the halo can be
better described in standard coordinates \cite{WEI}
\begin{equation}
\label{DS2N}
ds^2 = - B(r)dt^2 + A(r)dr^2 + r^2 d\Omega^2 \ \ .
\end{equation}
We have calculated the post--Newtonian solutions to the
equations of motion (\ref{SEQ1}, \ref{SEQ2}) as
\begin{equation}
\label{AGM}
\begin{array}{l}
B(r) =
{\displaystyle
1 - 2(\alpha - \sigma\delta)\frac{GM}{r} + ... }  \vspace{2mm}\\
A(r) =
{\displaystyle
1 + 2(\gamma - \sigma\delta)\frac{GM}{r} + ... }
\end{array}
\end{equation}
where
\begin{equation}
\label{abcd}
\alpha=1
\hspace{1cm}
\gamma=\frac{\omega+1}{\omega+2}
\hspace{1cm}
\delta=\frac{1}{\omega+2}  \ \ .
\vspace{1mm}
\end{equation}

Dynamical observations show that the velocity of objects away
from the disk of the galaxy remains constant for large
distances, suggesting that there is dark matter with
$M(r)\propto r\ $  \cite{MKT,RST}.
The measurements of the halo mass are obtained
from Kepler's third law $\ v^2=r g\ $ where $g$ is the
centripetal acceleration
${\displaystyle g=\ \frac{1}{2A(r)}\frac{dB(r)}{dr} }$.
For dark matter with variable masses Kepler's law gives
\begin{equation}
\label{3KL}
v^2=\ (\alpha-\sigma\delta)\
\frac{GM(r)}{r}\ +\ {\cal O}\left(\frac{GM}{r}\right)^2\ \ .
\end{equation}
Using eqs.(\ref{abcd}) and (\ref{OSB}) we find that the
coefficient of the linear term in (\ref{3KL}) is
\begin{equation}
\label{W2S}
1\ - \ 2.0\times 10^{-3}\ \leq\ \frac
{1+4\beta_I\beta_V}{1+4\beta_V^2}\leq\ 1\ +\  1.2\times 10^{-3}
\end{equation}
where we have used the bounds on $\beta_I\beta_V$ and $\beta_V$ which
will be given in sect.4 (see eqs.(\ref{BV2}, \ref{BIV})). Therefore,
the effect of variable masses for dark matter is rather
small. Since $M(r)\propto r$
corresponds to $\rho(r)\Phi^\sigma(r)\propto \frac{1}{r^2}$,
one could deduce the value of $\sigma$ from models of
galaxy formation for which the mass distribution $\rho(r)$ of
dark matter is known.

\subsection{Cosmological effects: Dark entropy}

We now consider a perfect fluid composed of dark matter particles
with variable masses. We can write the expression for the
energy--momentum tensor $T^{\mu\nu}$ and the particle number
current $N^\mu$ in the presence of a gravitational field as \cite{WEI}
\begin{equation}
\label{Tmn}
T^{\mu\nu} = p g^{\mu\nu} + (p+\rho) U^{\mu}U^{\nu}
\end{equation}
\begin{equation}
\label{Nm}
N^\mu = n U^\mu
\end{equation}
where $\ U^{\mu}\ $ is the local value of ${\ \displaystyle
\frac{dx^{\mu}}{d\tau}\ }$ for a comoving fluid element,
$\ g_{\mu\nu}U^{\mu}U^{\nu} = -1$ and $n$ is the particle
number density. Note that $p$ and $\rho$
are always defined as the pressure and energy density
measured in a local inertial frame comoving with the fluid,
and are therefore scalars.
$T^{\mu\nu}$ satisfies the energy conservation equation
(\ref{EMC}) while $N^\mu$ satisfies the particle number
conservation equation
\begin{equation}
\label{PNC}
N^\mu_{\ ;\mu} = 0 \ \ .
\end{equation}

The general cosmological solutions to the equations of motion
(\ref{SEQ1}, \ref{SEQ2}) in a Robertson--Walker frame
$ds^2 = - dt^2 + a^2(t)dx^2$
(compatible with the properties of a perfect fluid)
are \cite{FT27}
\begin{equation}
\label{CSOL}
\begin{array}{l}
a(t)\sim t^{\ p}, \hspace{1cm}
{\displaystyle
p = \frac{2(2\omega+3)-2(1-2\sigma)}
{3(2\omega+3)-2(1-2\sigma)+(1-2\sigma)^2}
} \vspace{2mm} \\
\Phi(t)\sim t^{\ q}, \hspace{1cm}
{\displaystyle
q = \frac{4(1-2\sigma)}
{3(2\omega+3)-2(1-2\sigma)+(1-2\sigma)^2}  } \ \ ,
\vspace{1mm}
\end{array}
\end{equation}
while the energy and particle number density conservation laws
(\ref{EMC},\ref{PNC}) expressed in the
Robertson--Walker metric give
\begin{equation}
\label{cm}
\frac{d}{dt}(\rho a^3) + p \frac{d}{dt}(a^3) =
\frac{1}{m}\frac{d m}{dt} (\rho-3p) a^3
\end{equation}
\begin{equation}
\label{cn}
\frac{d}{dt}(n a^3) = 0 \ \ .
\end{equation}

During the matter era ($T\ll m$) there is an entropy increase
(for variable masses) which can be computed by
comparing eq.(\ref{cm}) with the second law of Thermodynamics
\begin{equation}
\label{2LT}
dU + pdV = TdS \ \ .
\end{equation}

Taking, as an illustrative example, the case of a fluid of
non--relativistic monoatomic particles \footnote{It will become
clear that the physical consequences we will deduce do not rely
at all upon this assumption.} in thermal equilibrium at a
temperature $T$, the pressure and energy density are given by
\begin{equation}
\label{p}
\begin{array}{l}
p = n T \vspace{2mm}\\
\rho = n m + {\displaystyle \frac{3}{2} } n T
\end{array}
\end{equation}
from which we obtain an entropy production
\begin{equation}
\label{TdS}
TdS =\ N\left(1-\frac{3T}{2m}\right)dm\simeq\ N dm(\Phi) \ \ ,
\end{equation}
where $N=na^3$ is the number of dark matter particles in
thermal equilibrium. This equation can be integrated out
by using the fact that the particle number
conservation of non--relativistic matter species, with masses much
greater than their decoupling temperature $m\gg T_D$, imposes
the condition $m(t) T a(t)^2 =\ m(t_D) T_D a(t_D)^2\ $ \cite{MKT}.
The entropy increase from the time of equal matter and
radiation energy density to now, due to the variable
masses of dark matter, is given by
\begin{equation}
\label{TS}
\Delta S = \int_{t_{eq}}^{t_o} \frac{N dm}{T} \simeq
\frac{\sigma(1-2\sigma)}{2\omega+3 - (1-2\sigma)^2}\
\frac{N m(t_o)}{T_o} \equiv \ k(\omega,\sigma)\ \frac{N m(t_o)}{T_o} \ \ ,
\end{equation}
where $T_o$ is the invisible matter temperature at the present time $t_o$
and we have used the cosmological solutions (\ref{CSOL}).
Note that, {\em e.g.\ } for supersymmetric dark matter particles, their
mass is $m\simeq 1\ TeV$ and their particle number
$\ N \geq N_B \simeq 10^{-10} N_\gamma$. Then, using the fact that
its present temperature is not higher than that of the microwave background
radiation, we find
\begin{equation}
\label{DS}
\Delta S \geq \ k(\omega,\sigma)\ N_\gamma\ 10^{6} \ \ ,
\end{equation}
which is a huge number unless the prefactor $k(\omega,\sigma)$ is
very small. Note also that for the case $\ \sigma=0\ $ (JBD) and $\ \sigma=
\frac{1}{2}\ $ (GR), the entropy is exactly conserved \cite{MKT}.

This should be considered as a new source of entropy,
apart from the usual ones (cosmological phase transitions,
galaxy formation, star evolution, etc.).
It should be noticed that this increase is
actually an entropy exchange between the matter sector
and the gravitational sector through the coupling of
the scalar field. We are at present analyzing the physical effects
of such a dark entropy production.

\section{Phenomenological constraints}

As we have mentioned above, Damour {\em et al.} \cite{DGG}
constrained the parameters of the theory (\ref{SBD}) from radar
time--delay measurements, the age of the universe and the time
variation of $G$.
Visible baryonic matter dominates our solar system and therefore
can be constrained by PPN experiments. In particular, the Viking
experiments \cite{VIK} give bounds on the $\omega$ parameter,
$\ 2\omega+3 > 500$ \hspace{1.5mm} (95\% c.l.), which imply, see
eq.(\ref{OSB}),
\begin{equation}
\label{BV2}
\beta_V < 0.0224 \ \ .
\end{equation}
Bounds from cosmological observations
give $\ H_o t_o > 0.4\ $ \cite{WAF}, where $H_o$ is
the Hubble constant and $t_o$ is the age of the Universe. This
bound imposes a constraint on $\beta_I$ which is almost independent
of $\beta_V$, see eqs.(\ref{CSOL}, \ref{OSB}, \ref{BV2}), and gives
(see also \cite{DGG})
\begin{equation}
\label{BI2}
\beta_I < 0.674 \ \ .
\end{equation}

Since we are assuming that dark matter dominates
the cosmological evolution during the matter era, it will also be
constrained by bounds from primordial nucleosynthesis.
A recent analysis of these bounds was made in
Ref.\cite{FT26} for an ordinary JBD theory. Here, we
will generalize this result
to a theory with $\Phi$--dependent masses by using
the cosmological solutions (\ref{CSOL}) and the facts that
${\displaystyle \ G\sim \frac{1}{\Phi}\ }$ and \ $aT\sim$
constant. We first define the parameter
\begin{equation}
\label{XG}
\xi^2\equiv\frac{G_{rad}}{G_o}
= \left(\frac{T_{eq}}{T_o}\right)^{\frac{q}{p}} \ \ ,
\end{equation}
where $T_o$ and $T_{eq}$ are the cosmic background radiation
temperatures today and at the time of equal matter and radiation
energy densities, and $p$ and $q$ are the powers of time
evolution of the scale factor and scalar field, see eq.(\ref{CSOL}).

Primordial abundances of $\ D,\ ^3He$ and $\ ^7Li\ $
are quite sensitive to the baryon
to photon ratio $\eta_{10}$ ($\eta$ in units of $10^{-10}$).
Consistency, within two standard deviations, between observation
and GR--predictions forces $\eta_{10}$ to be in the range
$\ 2.6 \leq \eta_{10} \leq 4.0\ $ \cite{KAO}.
As will be seen shortly, the relevant bound for us is the lower
one, which comes from the upper observational limit on
$D +\ ^3He$. Of course, this bound must be slightly modified
when the theory under consideration is not exactly GR.
Concerning $^4He$, a good fit of the GR prediction for
$Y_p$ (mass fraction of primordial $^4He$) is \cite{KAO}
\begin{equation}
\label{YGR}
Y_p^{GR} = 0.228 + 0.010\ln\eta_{10} + 0.012 (N_{\nu}-3) + 0.185
\left(\frac{\tau_n-889.6}{889.6}\right)
\end{equation}
where $\tau_n$ is the neutron mean--life measured in seconds
$\ \tau_n = 889.6 \pm 5.8\ s  \ \ (2\sigma)\ $ \cite{WMA} and $N_{\nu}$
is the number of light neutrinos. Consistency between (\ref{YGR}),
the range of $\eta_{10}$, the experimental value of $\tau_n$
and the observational value
$\ Y_p^{obs} = 0.23 \pm 0.01   \ \ (2\sigma)\ $ \cite{MRE}
led in Ref.\cite{KAO} to bound the number of neutrinos to $N_{\nu}\leq3.4$\ .
We will compute the bounds on $\xi$ coming from nucleosynthesis.

It is easy to calculate the contribution to $Y_p$ coming from the
departure of ST theories from GR, since during the radiation era
all these theories behave like GR, except for a different
constant value of $G$.
More precisely, a variation $\delta G$ on Newton's constant
will be reflected in a variation $\delta Y_p$ of $\ ^4He\ $
production.  Actually, $Y_p$ depends only on the neutron--proton
abundance ratio $\ {\displaystyle x = \frac{N_n}{N_p}}\ $
at the time of nucleosynthesis $t_N$ \cite{MKT}
\begin{equation}
Y_p = \frac{2x}{1+x}|_{t_N} = \lambda\ \frac{2x}{1+x}|_{t_f}
\label{11}
\end{equation}
where
\begin{equation}
\lambda = {\displaystyle e^{-\frac{t_N - t_f}{\tau_n}}}\ \ .
\label{12}
\end{equation}
Here $t_f (T_f)$ represents the time (temperature) at which neutrons
'freeze out' from weak interactions and the exponential factor $\lambda$
accounts for the neutrons which decayed into protons between
$t_f$ and $t_N$ (this factor is very close to unity in all cases).
$T_f$ is determined by the competition between the expansion rate
($H \propto (NG)^{1/2}T^2$) and the weak interaction rate
($\Gamma_{weak} \propto G_F^2 T^5$). Therefore, $\delta T_f$ is
related to independent variations of $G$ and $N$. More precisely
\begin{equation}
\frac{\delta T_f}{T_f} = \frac{1}{6}\left(\frac{\delta G}{G}
+\frac{\delta N}{N}\right) \ \ ,
\label{14}
\end{equation}
where $N$ is the effective number of light species. We will suppose
that $N$ is fixed and compute the variation of $Y_p$ due only to
the variation of $G$ as \cite{FT26}
\begin{equation}
\delta Y_p = Y_p \left[\left(1 - \frac{Y_p}{2\lambda} \right)
\ln\left(\frac{2\lambda}{Y_p}-1\right) + \frac{3t_N-5t_f}
{\tau_n}\right] \frac{\ln 10}{3} \log \xi =\ 0.327 \log\xi
\label{YP}
\end{equation}
where we have used (\ref{14}), $\ \ln\xi\simeq\frac{\delta\xi}{\xi} =
\frac{1}{2}\frac{\delta G}{G}\ $ and substituted the central
value $\ Y_p^{obs} = 0.230\ $. Eq.(\ref{YP}) is
valid for {\em any} ST theory, as mentioned above.
The corresponding expression for $\xi$ should, however, be obtained
for each specific case. For a JBD theory with $\Phi$--dependent
masses, $\xi$ is related to $\omega$ through eq.(\ref{XG}), and
therefore, see eq.(\ref{CSOL}),
\begin{equation}
\label{LNX}
\frac{\omega+1+\sigma}{1-2\sigma}
= \frac{\ln(2\times 10^4\ \Omega h^2)}{2\ln\xi}\ \ .
\end{equation}

Comparing the theoretical prediction (\ref{YGR}, \ref{YP}) with
$Y_p^{obs}$, we obtain the desired bounds on $\xi$ and thus on
$\omega$ and $\sigma$. It is clear from (\ref{YGR}, \ref{YP},
\ref{LNX}) that the relevant bounds on $\omega > 0$ correspond
to lower bounds on $\eta_{10}$.
On the other hand, the lower bounds on $\eta_{10}$ come from
the upper observational limit on $(D +\ ^3He)_5$, the
primordial fraction of $D +\ ^3He$ in units of $10^{-5}$,
which also depends on $\xi$ \cite{DNS}
\begin{equation}
\log (D +\ ^3He)_5 = \ \log (D +\ ^3He)^{GR}_5
+ 2.3 \log\xi \ \ .
\label{21}
\end{equation}
Note that
$(D +\ ^3He)^{GR}_5$ is a known function of $\eta_{10}$
\cite{KAO}. Then, substituting the observational upper bound
$\ Y_{(D +\ ^3He)_5}^{obs} \leq 10.9   \ \ (2\sigma)\ $
\cite{KAO} in (\ref{21}), we obtain the lower bound on $\eta_{10}$
\begin{equation}
\ln\eta_{10} \ \geq \ \ln 2.60 + 3.7 \log\xi\ \ .
\label{E10}
\end{equation}
The constant term, $\ln 2.60$, corresponds to the  present GR
limit on $\eta_{10}$ for $N_\nu=3$. Future improvements on the
upper observational bound on $D +\ ^3He$ will increase this
number substantially \footnote{The upper bound on
$\eta_{10}$ comes from the upper bound on the
primordial abundance on $^7Li$ \cite{KAO}, which also depends on
$\xi$ and $\eta_{10}$. However, this is not relevant for the bounds
on positive $\omega$.}. Very conservative bounds on eq.(\ref{LNX})
from nucleosynthesis impose, for $\Omega h^2=0.25$,
\begin{equation}
\label{W250}
\frac{\omega+1+\sigma}{1-2\sigma}
>\ 380 \hspace{1cm} (95\%\ {\rm c.l.}) \ \ ,
\end{equation}
which constrains the parameters of our theory as
\begin{equation}
\label{BIV}
\beta_I\beta_V < 3\times 10^{-4} \ \ .
\end{equation}
Suppose that string theory or any other fundamental theory of gravity
fixes the coupling $\beta_I\sim {\cal O}(1)$. Then the constraint
(\ref{BIV}) would dramatically improve the bound on $\omega$ to
$2\omega+3 > 2\times 10^6$.

In the next section we will introduce a general scalar--tensor
theory with variable gauge couplings, as predicted by string
theory, and give even stronger constraints on the parameters of
the theory.

\section{String theory}

String theory gives definite predictions on the couplings of
dilaton and moduli fields to the matter sector, which depend on
the compactification and symmetry breaking mechanism. However,
it also predicts a scalar coupling of the dilaton to the
Yang--Mills sector (and in particular to the electromagnetic sector)
which has not been previously considered to constrain generalized JBD
theories, such as that of eq.(\ref{SMD}). This coupling, to be added to
(\ref{SMD}), has the form
\begin{equation}
\label{F2}
-\frac{1}{4}e^{-\phi} F_{\mu\nu}F^{\mu\nu}\;\;,
\end{equation}
so $\alpha_{e.m.}\propto e^{-\phi}$ (at tree level).
In consequence, a variation of the electromagnetic coupling
constant is related
to the corresponding variation of Newton's constant by
$\ \frac{\delta\alpha_{e.m.}}{\alpha_{e.m.}}=\lambda\frac{\delta G}{G}\ $,
where $\ \lambda=(2\omega+3)^{1/2}$, see eq.(\ref{OSB}).
There are bounds on the variation of $\alpha_{e.m.}$
which are extraordinarily strong \cite{DYS}.
For example, Dyson gives a bound on $\ \delta\alpha_{e.m.}\ $
from the nuclear stability of the $\beta$--isotopes
$\ ^{187}_{75}Re$ and  $\ ^{187}_{76}Os$,
$\ |\frac{\delta\alpha_{e.m.}}{\alpha_{e.m.}}|<\ 2.5\times 10^{-5}\ $
since the formation of the Earth.
This bound translates into the parameters of our theory as
\begin{equation}
\label{STI}
\beta_I\beta_V<\ 1.7\times 10^{-8} \ \ .
\end{equation}
For a scenario in which all the masses have the {\em same}
$\Phi$--dependence ($\beta_I=\beta_V$), this imposes a
strong bound on $\omega$
\begin{equation}
\label{STO}
|2\omega+3| >\ 1.4\times 10^8 \ \ .
\end{equation}
For other values of $\beta_I$, the bounds are even stronger.
{\em E.g.} for $\ \beta_I\simeq 0.5$  (see eq.(\ref{BI2}))
we obtain
\begin{equation}
\label{STB}
|2\omega+3| > \ 3.4\times 10^{15} \ \ .
\end{equation}
As we can see, these bounds are much greater than any other
bound from nucleosynthesis or post--Newtonian experiments.
Other bounds on the constancy of $\alpha_{e.m.}$ (e.g. from Oklo
reactor data \cite{DYS}) give constraints on $\omega$ that
can be even stronger.
The previous results are modified when one takes into account that
the electromagnetic constant measured at low energy is corrected with
respect to the tree level one by the renormalization group contributions,
but this does not affect the bounds substantially.

We will conclude this section with several comments. First, the
previous bounds on $\omega$
are general, in the sense that they will be always hierarchically larger than
the usual ones (i.e. post--Newtonian and nucleosynthesis bounds) for
any considered scenario. Second, these bounds are so strong that they
virtually exclude any possibility of having a phenomenologically
viable generalized JBD theory (e.g. that of eq.(\ref{SMD})) based
on the dilaton field of superstrings. This means that a non--trivial
potential (presumably from non--perturbative effects, like gaugino
condensation \cite{GAU}) fixing the value of the dilaton is mandatory, thus
leading to a theory (with $<\phi>=$constant) in which the dilaton does
not play any more the role of a JBD scalar. Notice however that this is
required only during the matter dominated era, for which the previous
bounds have been obtained. For example, if the potential which fixes
the dilaton develops after an inflationary epoch, which seems reasonable,
an extended inflationary scenario based on $\phi$ can occur, see
\cite{FT37}. In any case
this does not mean that at present the only viable theory of gravity
from strings is ordinary GR. As mentioned above, in the effective four
dimensional theory from strings besides the dilaton $\phi$ there are
other neutral scalar fields called {\em moduli}, which usually parametrize
the size and shape of the compactified space, sharing many properties
with $\phi$. In particular, they possess a flat potential at all orders
in string perturbation theory and have couplings with gravity and matter
fields, whose masses become in this way moduli--dependent (see e.g.
ref.\cite{MASS}). However the moduli are {\em not} coupled to the
Yang--Mills sector, as the dilaton in eq.(\ref{F2})\footnote{Actually,
it has been recently shown \cite{THRES} that, depending on the
compactification
scheme, some of the moduli can be coupled to $F_{\mu\nu}F^{\mu\nu}$
through one loop corrections. In any case, there is only one combination
of the dilaton with the moduli playing the role of the electromagnetic
coupling constant.}, so the restrictions found in this section do not apply
for a generalized JBD theory based on these fields. On the other
hand, it is perfectly conceivable that the non--perturbative
potential that fixes the electromagnetic coupling constant still leaves
flat directions along certain combinations of moduli. For instance,
the gaugino condensation effects do not usually involve all the moduli
of the theory \cite{THRES}. Then, these fields would really play the
role of JBD scalars with the usual bounds on the corresponding value
of $\omega$.

\section{Conclusions and Outlook}

We have studied new physical properties and phenomenological
constraints of a generalized ST theory of gravity with
$\Phi$--dependent masses. Under conformal redefinitions of the
metric it is always possible to make visible masses constant.
We thus choose the frame in which the units of measure are constant.
Nevertheless, physical predictions do not depend on the
conformal frame. In consequence, theories where all
the masses have the {\em same} $\Phi$--dependence are equivalent
to ordinary ST theories.

We have investigated a scenario with two types of
$\Phi$--dependent masses (thus violating the weak Equivalence
Principle) which could correspond to visible and dark matter
sectors. This possibility can arise in a natural way in the
low--energy effective theories from superstrings and leads to
new and interesting physics. We have constructed a
perfect fluid formalism for the dark matter sector
(which in the physical conformal frame has variable masses),
finding a non--adiabatic cosmological expansion during the matter era,
as a consequence of the non--conservation of the matter energy--momentum
tensor. The latter can be physically interpreted as an energy exchange
between $\Phi$ and the matter sector. It is actually a
dissipative effect, which produces a significant entropy increase
during the matter era. The implications of such a 'dark entropy'
are currently under investigation.

We have analyzed some phenomenological constraints on this theory
assuming that dark matter dominates the cosmological evolution
during the matter era, while visible matter dominates our solar
system. Post--Newtonian bounds constrain only the visible sector
as in ordinary ST theories. However, nucleosynthesis bounds are
sensitive to both sectors, since the universe expansion, during
the matter era, is essentially governed by dark matter. It turns
out that usual bounds on the visible sector would be
dramatically improved, provided we had some (theoretical)
knowledge on the $\Phi$--dependence of dark matter masses.

Another appealing new physical effect of this kind of
scenario is that dark matter with $\Phi$--dependent masses
would have, in principle, a measurable effect on the dynamical motion
of the halo of spiral galaxies. From post--Newtonian solutions of
JBD theory with $\Phi$-dependent masses, we have obtained the effect of
variable masses on the centripetal acceleration of objects away from
the disk of the galaxy.

In the case of string effective theories there is also a
$\Phi$--dependent gauge coupling constant. The bounds on the
constancy of gauge couplings are so strong that we obtain bounds
on the parameters of our theory which are many orders of
magnitude stronger than the previous bounds. These bounds do not apply,
however, for other JBD--like fields of the theory (the moduli).

Finally, we have used in most examples a very simple coupling of
the scalar field to masses in which the $\Phi$--dependence is
power--law. This assumption allows for an arbitrary
constant $\omega$ parameter in the corresponding JBD theory.
However, the $\Phi$--dependence could be any arbitrary function,
which would then correspond to general $\omega(\Phi)$
functions. In that case, our conformal analysis can be
straightforwardly generalized to scalar--tensor theories of gravity.

\nsect*{Note added}

After completion of this work, we became aware of Ref.\cite{DG}
where bounds from nucleosynthesis, similar to those presented
in sect.4 of this paper, were performed. Our results are
slightly stronger than theirs since they are based on updated
observational and experimental bounds.

\end{document}